\documentclass[11pt,aps,prl,linspace=1 superscriptaddress]{revtex4-1}
\usepackage{amsmath}
\usepackage{graphicx}
\usepackage{xcolor}
\usepackage{gensymb}
\usepackage{hyperref}
\usepackage{upgreek}
\usepackage{braket}
\usepackage[pagewise]{lineno}
\usepackage{float}

\begin{document}
\title{Large Area Automated Characterisation of Chemical Vapour Deposition Grown Monolayer Transition Metal Dichalcogenide Through Photoluminescence Imaging}
\author{T. Severs Millard$^a$, A. Genco$^a$, E. M. Alexeev$^a$, S. Randerson$^a$, S. Ahn$^b$, A. Jang$^b$, H. S. Shin$^b$ and A. I. Tartakovskii$^a$}
\affiliation{$^a$ Department of Physics and Astronomy, University of Sheffield, Sheffield, UK}
\affiliation{$^b$ Department of Energy Engineering and Department of Chemistry, Ulsan National Institute of Science and Technology (UNIST), Ulsan, South Korea. }

\date{\today}

\begin{abstract}
Chemical vapour deposition (CVD) growth is capable of producing multiple single crystal islands of atomically thin transition metal dichalcogenides (TMDs) over large area substrates, with potential control of their morphology, lateral size, and epitaxial alignment to substrates with hexagonal symmetry. Subsequent merging of perfectly epitaxial domains can lead to single-crystal monolayer sheets -- a step towards scalable production of high quality TMDs. For CVD growth to be effectively harnessed for such production it is necessary to be able to rapidly assess the quality of material across entire large area substrates, characterising the properties of islands and allowing causality to be found with growth parameters. To date characterisation has been limited to sub 0.1 mm$^2$ areas, where the properties measured are not necessarily representative of an entire sample. Here, we apply photoluminescence (PL) imaging and computer vision techniques to create an automated analysis for large area samples of semiconducting TMDs, measuring the properties of island size, density of islands, relative PL intensity and homogeneity, and orientation of triangular domains. The analysis is applied to 20x magnification optical microscopy images that completely map samples of WSe$_2$ on hBN, 5.0 mm x 5.0 mm in size, and MoSe$_2$-WS$_2$ on SiO$_2$/Si, 11.2 mm x 5.8 mm in size. For the latter sample 100,245 objects were identified and their properties measured, with an orientation extracted from 27,779 objects that displayed a triangular morphology. In the substrates studied, two prevailing orientations of epitaxial growth were observed in WSe$_2$ grown on hBN and four predominant orientations were observed in MoSe$_2$, initially grown on c-plane sapphire. The proposed analysis will greatly reduce the time needed to study freshly synthesised material over large area substrates and provide feedback to optimise growth conditions, advancing techniques to produce high quality TMD monolayer sheets for commercial applications.

\end{abstract}

\maketitle

Van der Waals layered crystals such as TMDs, hexagonal boron nitride (hBN) and graphene are defined by their strong covalent bonds in plane and weak interlayer forces \cite{geim2013van}. These characteristics allow for individual atomically thin layers to be easily removed from the bulk crystal, and for single layers to be brought together to build vertical heterostructures that display promising new properties \cite{novoselov2005two,geim2013van,withers2015light}. As TMDs are exfoliated down to monolayer thickness their electronic band structure is altered due to dimensional confinement, causing the materials to become direct bandgap semiconductors that display photoluminescence (PL) under optical excitation\cite{mak2010atomically,splendiani2010emerging}. MoSe$_2$, WSe$_2$, MoS$_2$ and WS$_2$ are documented as the most promising and widely studied TMD monolayers, emitting bright PL while remaining stable in air, at room temperature \cite{tonndorf2013photoluminescence}. Such traits lend these materials to applications in optoelectronic devices \cite{pospischil2016optoelectronic} such as LEDs \cite{withers2015light,ross2014electrically}, photovoltaic cells \cite{pospischil2014solar,wu2019highly}, photodetectors \cite{huo2018recent} and single photo-emitters \cite{koperski2015single,palacios2018atomically}. Their optical properties have attracted further attention due to the possibility for straightforward coupling to spin and valley degrees of freedom \cite{XuNatRevMat2016}. The extreme thinness and mechanical stability of these materials provide the potential for flexible and transparent devices \cite{akinwande2014two}. However, to realise such devices, the goal of scalable and controlled production of mono- and multilayers must first be achieved, an objective which has been of great interest for commercial application ever since TMDs were rediscovered as atomically thin materials \cite{mak2010atomically,splendiani2010emerging}.

The original method of monolayer production, micromechanical exfoliation \cite{novoselov2005two}, produces high quality two-dimensional (2D) flakes, but with a low yield and random nature it has an inherent inability to be scaled up. This has lead to a catalogue of methods being developed including other top-down approaches such as liquid phase exfoliation \cite{backes2016guidelines} and various forms of thinning \cite{castellanos2012laser,varghese2017topography,huang2013innovative}, as well as bottom-up routes such as wet-chemical synthesis \cite{samadi2018group} and physical vapour deposition \cite{wasa2012handbook,muratore2014continuous}. A further technique, chemical vapour deposition (CVD), has the capability to grow multiple monolayer islands across large area substrates and has been earmarked as the process that will deliver scalable production \cite{shi2015recent}.

Continuous lateral growth of these CVD islands leads to the in-plane merging of single-crystal monolayers and eventually the formation of monolayer sheets. A problem encountered is that if randomly orientated, or irregularly shaped islands are allowed to coalesce, grain boundaries  are formed at every merger of unaligned edges, which may negatively affect the optical and electrical characteristics of the sheet \cite{najmaei2014electrical,ly2016misorientation}. The density of grain boundaries can be reduced through epitaxial growth of the TMD islands on substrates with hexagonal symmetry, such as hBN \cite{okada2014direct,yan2015direct,yu2017precisely}, mica \cite{ji2013epitaxialSEM}, c-plane sapphire \cite{dumcenco2015large,zhang2018diffusion} and graphene \cite{bianco2015direct}. On these substrates, van der Waals interlayer interaction promotes the relative alignment despite the, often large, lattice mismatch \cite{nakanishi2019atomic}. In CVD the morphology of islands can be controlled to a degree by altering the local ratio of the transition metal and chalcogen atoms around a nucleation point. Typically the synthesised islands take on a hexagonal or equilateral triangle shape \cite{yang2017effective,wang2014shape}. The regular and reproducible triangular morphology is of particular interest with respect to the production of continuous single-crystal films. On the majority of the crystalline substrates mentioned, hBN \cite{okada2014direct,yan2015direct,yu2017precisely}, mica \cite{ji2013epitaxialSEM} and graphene \cite{bianco2015direct}, two well defined groups of triangular islands rotated at 60$^o$ relative to one another are observed. The subsequent merging of these two groups results in either perfect stitching between objects at the same orientation, or the formation of mirror grain boundaries between those of opposite. On c-plane sapphire two more orientations were observed with a much lower probability, both at a 30$^o$ rotation from the first set \cite{dumcenco2015large}. 

However, epitaxial growth is not a well studied phenomenon for any TMD and substrate combination, and it is generally only assessed manually across sub 0.1 mm$^2$ areas, with analysis containing just hundreds of triangular crystals at most. Such small sample sizes are not necessarily representative of the CVD growth across an entire substrate, as all island properties, both physical and optical, vary with spatial position. If this ordered growth can be analysed across a larger area and a causality established with growth parameters, the gained feedback could be used for production of monolayer sheets with minimal grain boundaries, potentially realising commercial production of devices. Further requirement for regular orientation of islands arises from potential uses in van der Waals heterostructures, where a relative twist between adjacent layers is being actively researched as a new degree of freedom \cite{VanDerZande2014,Liu2014,Kunstmann2018,alexeev2019resonantly,JinNature2019,
SeylerNature2019,TranNature2019}.

Although the mechanisms that bring about the physical properties of CVD grown material are believed to be understood \cite{samadi2018group}, to date, there are no documented methods to provide feedback on an entire sample for optimisation of the desired properties such as, island size and density, quality of PL emission and orientation of individual islands. If the quality of CVD grown TMD monolayers is to improve and high grade sheets and van der Waals heterostructures produced, it is first necessary to be able to reliably characterise entire large area samples. 

Computer vision has been used in the field of 2D materials, predominantly as a tool for automatic mono- and multilayer identification of exfoliated material across large areas \cite{jessen2018quantitative,lee2018highly} -- freeing up time taken searching for flakes. The technique of PL imaging has been developed in parallel, as a fast and innovative type of optical imaging, that can be simply applied to a standard optical microscope for wide-field and fast capture of fluorescent monolayer material \cite{alexeev2017imaging}.

Here, these two techniques have been applied to create an automated analysis capable of characterising substrates on which CVD synthesis has been carried out. 
The size, orientation, relative intensity and homogeneity of PL emission are analysed using image processing software for each TMD island on substrates up to 65 mm$^2$ is size, containing more than 100,000 individual islands of one or two TMD materials.
The analysis is demonstrated in application to two samples of WSe$_2$ grown directly onto hBN (sample 1 and 2, both 5.0 mm x 5.0 mm) and a third sample (sample 3) containing MoSe$_2$ and WS$_2$ islands on SiO$_2$/Si (11.2 mm x 5.8 mm), where the WS$_2$ was grown directly onto the substrate while the MoSe$_2$ was grown on c-plane sapphire and subsequently transferred. The reported characterisation and analysis can be extended to any combination of TMDs and substrate where PL is emitted at room temperature. For WSe$_2$ monolayers grown directly on hBN, we observe two equally preferential orientations, while MoSe$_2$ samples originally grown on c-plane sapphire and subsequently transferred onto SiO$_2$/Si substrate demonstrate four prevalent preferential orientations.


\renewcommand{\baselinestretch}{1.2}

\begin{figure*}[b!]
	\centering
	\includegraphics[width=1\linewidth]{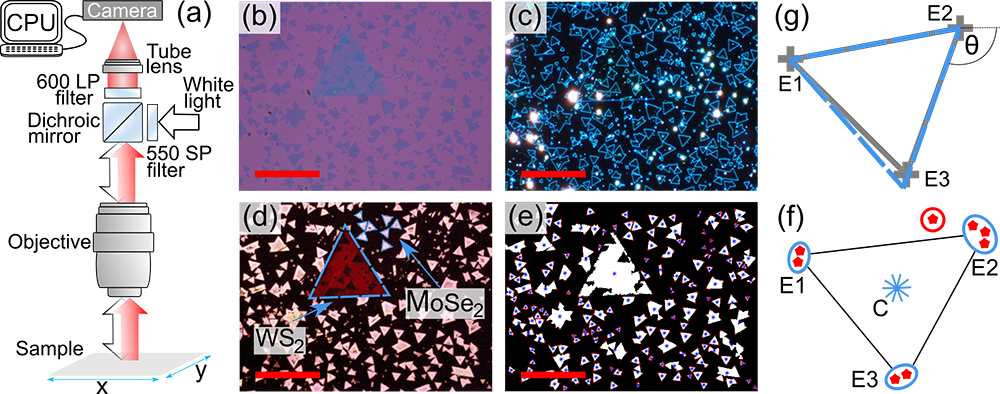}
	\caption{\textbf{Imaging and image processing of CVD grown TMD monolayers.} (a) 
(a) A schematic showing how PL emission is isolated from the collected light in a typical BF microscope set-up using a 550 nm short-pass filter, 550 nm long-pass dichoric mirror and 600 nm long-pass filter combination. (b-d) A comparison between images at 50x magnification obtained using BF (b), DF (c) and PL (d). In (d) the two materials are outlined in light blue with MoSe$_2$ appearing as a light pink colour and WS$_2$ as a dark red. The image was taken with a 10 s exposure time and 9.6x analogue gain. The red scale bars in each image are 50 $\mu$m. The binary image resulting from the image processing of (d) is shown in (e), with each individual object highlighted by their central point (centroid) shown by a blue marker, and their extrema shown by  red dots. (f) Illustration of how the applied image analysis program attempts to find a triangular outline of a single isolated island by combining or discarding the extrema, resulting in three clear combined points, E1-3. See further details in text and Supplementary Information (Supplementary Note 1). (g) A visual representation of how the program calculates the orientation of each island from the positions of the combined extrema, E1-3. The light blue equilateral triangle is fitted to each combinations of two points and the angle, $\theta$, along with an uncertainty, is calculated.
	\label{fig:a}}
\end{figure*}

\noindent \textbf{Results.}
\\\textbf{Photoluminescence imaging and image recognition techniques.}  
We use computer vision functions from the MATLAB Image Processing Toolbox \cite{matlabImage} to analyse the high-resolution PL image of the entire sample on a large substrate of a few 10s of mm$^2$, acquired by manually stitching the images taken with a 20x magnification microscope objective. The monolayer TMD islands were identified through their PL as the only luminescent objects on the studied substrates, and their size and shape were extracted using MATLAB shape recognition functions. Only monolayer islands with the shape close to equilateral triangle were used for orientation analysis, while monolayer islands with all morphologies were used for size, mean PL intensity, and material coverage analysis, as described below in this section, and further detailed in Methods and in Supplementary information.

Fig. \ref{fig:a}. (a) shows a schematic diagram of how a dichroic mirror can be used to isolate PL emission in a generic microscope set-up, reflecting the visible light while transmitting in the near infrared \cite{alexeev2017imaging} (see details in methods). A comparison between images of monolayers at 50x magnification in bright-field (BF), dark-field (DF) and PL modes is shown in Fig. \ref{fig:a}. (b-d), respectively, for the same area of the MoSe$_2$-WS$_2$ sample on SiO$_2$/Si (referred to as sample 3 below). In the BF image objects are distinguishable from the purple substrate, although contrast is low and the two materials are difficult to discriminate. In the DF image edges are well defined, but other features on the sample surface, such as dust, appear bright -- masking monolayers. On top of this, the edges of both materials appear as the same colour and objects of WS$_2$ are only recognisable from those of MoSe$_2$ by their size under careful analysis. The PL image shows the greatest contrast between the substrate and materials as only the latter have the potential to display PL, the presence of which confirms their monolayer thickness. The material dependent wavelength of the emitted light at room temperature allows multiple TMDs to be easily identified in a single image and analysed separately \cite{alexeev2017imaging}, with PL peaked at 785 nm for MoSe$_2$, 750 nm for WSe$_2$, 675 nm for MoS$_2$, and 630 nm for WS$_2$, where wavelengths above 700 nm appear as false colours. The boundaries of all isolated islands are clear and the dominant triangular morphology is evident \cite{wang2014shape}. Compared to DF and, to a degree, BF, unwanted contaminants such as dust particles have little effect on the quality of a PL image. 

In order to measure properties of the islands, size, orientation, density and relative PL emission, it is necessary to first convert the PL image into a binary form, with pixel values set to either on (1), representing PL emission and thus a monolayer, or off (0), for areas of substrate or thicker material. This is achieved through colour thresholding to isolate all red and light pink coloured objects in an image, as shown in Fig. \ref{fig:a}. (e). In this form, image processing functions, taken from MATLAB's image processing toolbox \cite{matlabImage}, can be applied to improve the quality of analysis (see details in Supplementary Note I). The full colour 8-bit image is also converted to grayscale, where each pixel is represented by a single value between 0 and 255. This pixel value is calculated from the channels of the given format that define its colour before the conversion, in this case red, green and blue (RGB). The properties of islands are measured from both binary and grayscale images. Size and extrema, which are used to find orientation, are taken solely from the binary image, and mean pixel value and standard deviation (STD) of pixel value are extracted from the masked grayscale image. The latter two properties provide a relative measure of the average brightness and homogeneity of PL emission across an objects surface.

To measure the orientation of each object the assumption is made that all islands are equilateral triangles, the most common shape seen in Fig. \ref{fig:a}. (b-e). As mentioned in the introduction, this is a well documented morphology whose growth mechanism and parameters are reasonably well understood \cite{wang2014shape,govind2016generalized,you2018synthesis}. The positions of the extrema, shown in red in Fig. \ref{fig:a}. (e, f), are called from the `regionprops' MATLAB function, returning the furthest eight points on an objects surface from its centroid. These are averaged together based on proximity to one another and size of the given island. Only the vertexes that are defined by at least two of the extrema returned by the function are considered for the analysis, thus ensuring an unambiguous identification. 
If an island is found to have anything but three vertexes it is not a triangle and is discarded. 
Fig. \ref{fig:a}. (f) demonstrates this step, where the positions of the extrema are marked by red stars and the position of the centroid is marked by a blue asterisk. Those extrema contained in a blue oval would be combined to an average position, and the one circled in red would be discarded. An equilateral triangle is then fitted to each combination of two vertexes and an orientation found. Fig. \ref{fig:a}. (g) shows this step, where the light blue equilateral triangle is fitted to the points E1 and E2, and $\theta$ is the calculated orientation. Typically an island grown via CVD is not perfectly equilateral, and so fitting an equilateral triangle to each side gives us a range of orientations which can be used to find an uncertainty. This, in turn, can be applied to filter out islands displaying incorrect morphology by discarding any measurement with a percentage uncertainty greater than 30\% of the mean value for the data set. A more detailed explanation of the image processing and the steps of the analysis can be found in the Supplementary Note I.

\begin{figure*}
	\centering
	\includegraphics[width=1\linewidth]{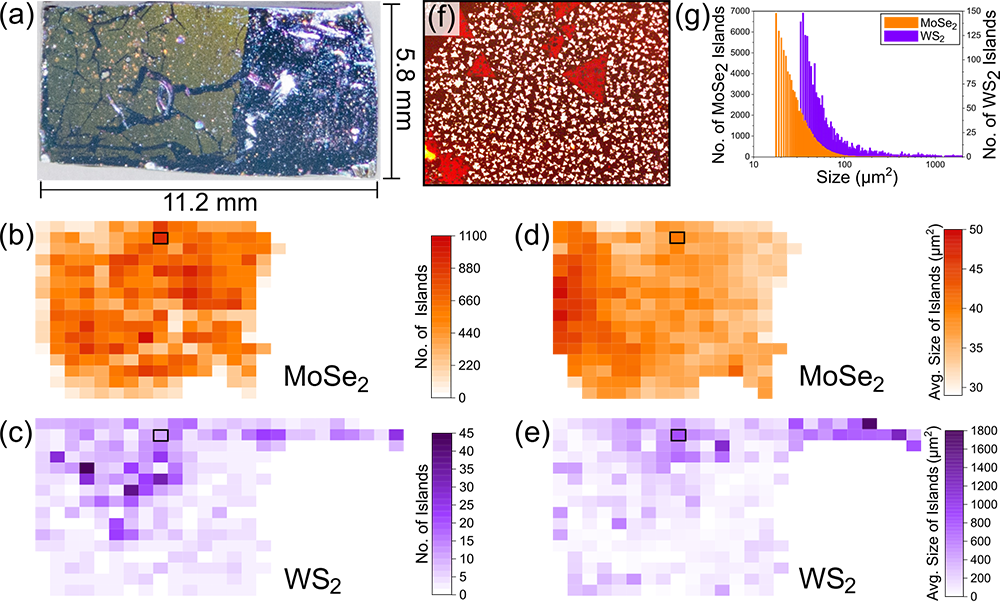}
	\caption{\textbf{The size and density of MoSe$_2$ and WS$_2$ islands across a large area CVD grown substrate.} (a) A macroscopic image of the MoSe$_2$-WS$_2$ sample with dimensions of 11.2 mm x 5.8 mm. Heatmaps (b) and (c) show the density of isolated islands of MoSe$_2$ and WS$_2$ across the same sample, respectively. Heatmaps (d) and (e) display the average size of islands for MoSe$_2$ and WS$_2$. Each pixel in the heatmaps represents an average value for the given property across all objects contained in a single 20x magnification image, 373 $\mu$m x 497 $\mu$m in area. The white spaces to the right of the maps show empty areas of substrate. The black box on each map highlights the area shown in (f) to provide an example of the typical growth of materials and to better understand the scale on the maps. The histogram (g) shows the distribution of sizes of isolated objects for each material. The offset for both materials is a product of removing small objects from the analysis. The results shown here are from 97,672 isolated MoSe$_2$ islands and 2,573 of WS$_2$.
	\label{fig:b}}
\end{figure*}

Fig. \ref{fig:a}. (d) and (e) show some limitations of the analysis. If two single domain islands grow into one another the proposed analysis fails to identify the position of the grain boundaries from the PL image. Thus, in the binary image, the two islands will be treated as a single object whose orientation data will be discarded. However, across an entire substrate of dimensions 11.2 mm x 5.8 mm, seen in Fig. \ref{fig:b}. (a), an orientation could be measured for 28.2\% of the objects -- proving to be sufficient to gauge the level and strictness of aligned growth. It is important to note that only the orientation analysis discards data, all other properties such as size, mean pixel value, STD of pixel value and island density, are taken from all monolayer area, regardless of shape. This orientation measurement method could be easily adapted for a hexagonal morphology.

\renewcommand{\baselinestretch}{1.2}

\renewcommand{\baselinestretch}{1.2} 
\begin{figure*} [b!]
	\centering
	\includegraphics[width=1\linewidth]{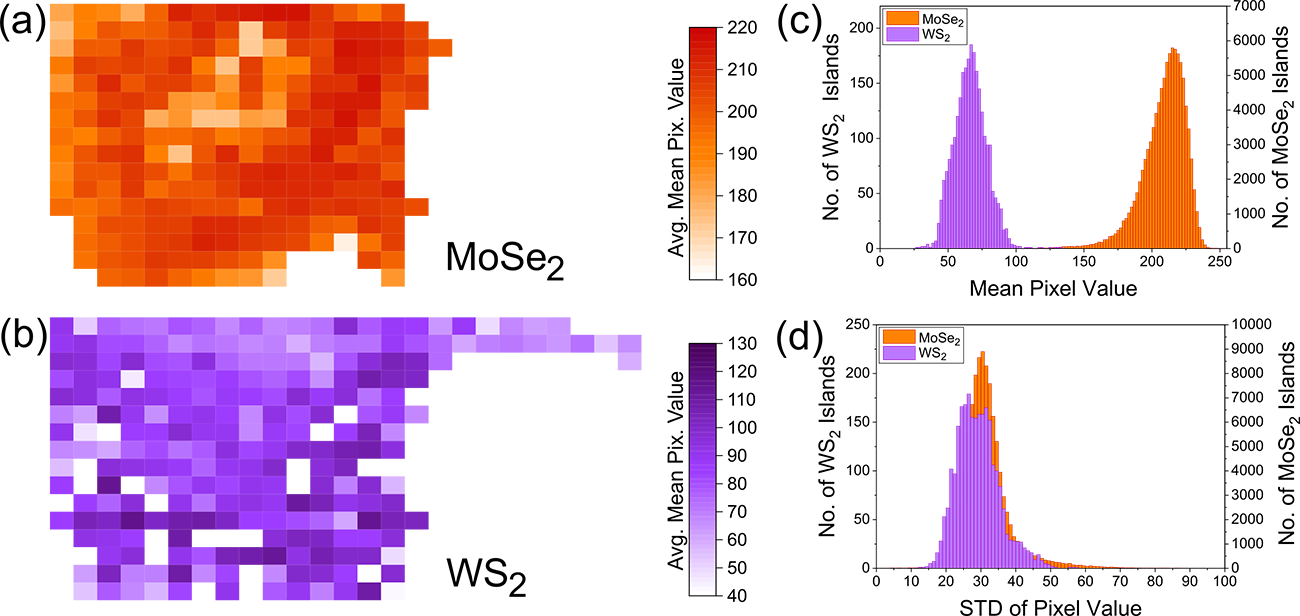}
	\caption{\textbf{Assessment of the relative PL emission of monolayers across the MoSe$_2$-WS$_2$ sample measured through the pixel value of objects.} Heatmaps displaying an average of the mean pixel value for every object contained in each 20x magnification image that stitch together to completely map the sample of MoSe$_2$-WS$_2$ on SiO$_2$/Si are shown in (a) for MoSe$_2$, and (b) for WS$_2$. The distribution (c) contains the average pixel intensity for all 100,245 objects identified by the program, providing an insight into the variation of PL between islands. The distribution of the STD of pixel value across the surface of the same objects is shown in (d), displaying variation in PL emission on an island by island basis.
	\label{fig:c}}
\end{figure*}

\textbf{Physical and optical properties of individual islands.}
A macroscopic image of the MoSe$_2$-WS$_2$ sample on SiO$_2$/Si can be seen in Fig. \ref{fig:b}. (a), showing the presence of material but no details of individual islands. The entire sample was manually mapped in PL imaging mode at 20x magnification with 10 s exposure time and 9.6x analogue gain, as described in Methods, and the set of image processing tools outlined above were applied to each frame (373 $\mu$m x 497 $\mu$m). Details of the fabrication of this sample can also be seen in Methods.

Heatmaps displaying the number of isolated MoSe$_2$ and WS$_2$ objects contained in each image, represented by a single tile on the map, are shown in Fig. \ref{fig:b}. (b) and (c). The size of all islands in each image are averaged and represented by a tile on the heatmaps of Fig \ref{fig:b}. (d) and (e). It can be seen that MoSe$_2$ islands are much more homogeneous in size across the surface of the substrate, although flakes do get larger closer to the middle of the left edge. WS$_2$ vary more in size, and the areas containing the largest islands appear to roughly align with the sections of highest number density. What is quite clear from the heatmaps, and in agreement with the reference image Fig. \ref{fig:b}. (f), whose position on the maps is outlined by a black box, is that MoSe$_2$ islands are small and densely packed together while the WS$_2$ islands are much larger and have a lower nucleation density. MoSe$_2$ is seen across two thirds of the substrate from left to right while WS$_2$ has mostly developed along the top. 

The histogram, Fig. \ref{fig:b}. (g), displays a distribution of size with a logarithmic scale for all objects on the sample. This shows that the WS$_2$ tails off at much larger areas and that the smallest islands are dominant in number for both materials. Both distributions are offset from 0 since objects, less than 200 pixels (19.3 $\mu$m$^2$) in size for MoSe$_2$ and 300 pixels (28.9 $\mu$m$^2$) for WS$_2$, were omitted in the image analysis. 

If a monolayer sheet completely covering the substrate is considered 100\% coverage, the identified 97,672 isolated MoSe$_2$ objects cover 6.90\% of the substrate and 2,573 WS$_2$ objects cover 0.77\%. If the 39.4\% of the substrate that is completely empty of any material, shown by the white area at the bottom right of the heatmaps, is omitted from the calculation, then the total coverage is 11.0\% and 1.24\% for MoSe$_2$ and WS$_2$, respectively. These coverage values are slightly underestimated as they do not include small objects removed during the image processing. 

The optical properties of a sample can be displayed in much the same way as the size and density information, to reveal how PL emission changes across the surface of a sample. The heatmaps in Fig. \ref{fig:c} show how the average of the mean pixel value for each object in an image changes with position on the sample for MoSe$_2$ (a) and WS$_2$ (b). In the PL imaging mode the mean pixel value for an object is directly related to the strength of that object PL emission, the camera settings used to capture the image and the optical excitation power. 
Relative comparison between the properties of individual material within a heatmap is possible, as all images were acquired with identical gain and exposure settings, and with the same excitation power. 
For MoSe$_2$, Fig. \ref{fig:c}. (a) shows clear gradients between areas of the highest and lowest PL emission, while WS$_2$ in (b) shows a more random nature with respect to spatial position. The distributions seen in Fig. \ref{fig:c}. (c) can shed some light on the variation in the average PL emission of individual islands across the sample through their full width half maximum (FWHM), measured as 28 for both materials. The standard deviation (STD) from the mean pixel value of an object tells us how much these values vary across its surface and therefore how homogeneous its PL emission is. As a distribution, Fig. \ref{fig:c}. (d), STD allows the consistency of the optical emission to be seen on an island by island basis, across an entire substrate. Here, the PL emission homogeneity of WS$_2$ islands may be slightly compromised by the quenching that occurs where heterostructures are formed with the smaller islands of MoSe$_2$ \cite{alexeev2017imaging}.

\renewcommand{\baselinestretch}{1.2} 
\begin{figure*} 
	\centering
	\includegraphics[width=0.95\linewidth]{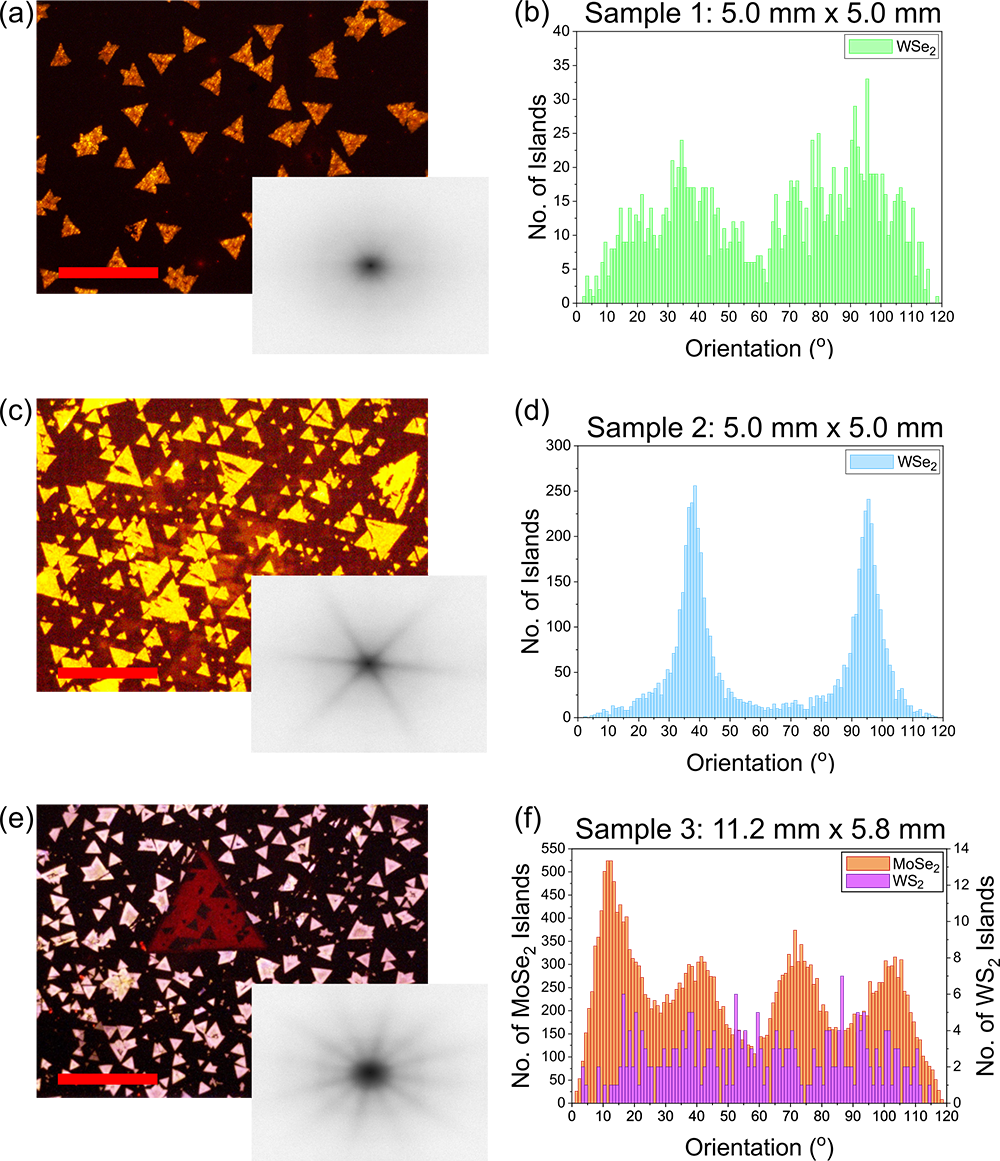}

	\caption{\textbf{Orientation analysis to determine the level of epitaxial growth in three CVD samples.} (a) An example PL image at 50x magnification of sample 1 containing WSe$_2$ grown on a 5.0 mm x 5.0 mm square of hBN, with no obvious signs of alignment. The FFT inset to the image shows data from the whole sample. (b) shows the results of the orientation analysis for this sample as described in text. (c) A 50x PL image for sample 2, WSe$_2$ grown on hBN, showing high degree of alignment. The inset shows the FFT uncovering the pattern and periodicity across the entire sample, while (d) shows the histogram of orientation for all objects of WSe$_2$ which could be measured by the applied image processing program. (e) A 50x PL image of sample 3 discussed in Figs.1-3, showing MoSe$_2$ (pink) and WS$_2$ (red) monolayers. The FFT in the inset shows results for MoSe$_2$ monolayers showing 12 prevailing orientations of the edges of the triangular islands at 30$^o$ to each other.  (f) Shows the orientation distribution for both materials in (e), showing epitaxial arrangement for MoSe$_2$ islands and random distributions for WS$_2$. The red scale bars in (a), (c) and (e) are 50 $\mu$m.
	\label{fig:d}}
\end{figure*}

\textbf{Analysis of crystal orientation over 25 to 65 mm$^2$ substrates.} The strictness of epitaxial growth to the hexagonal lattices of their substrates has been assessed for the three samples introduced earlier. Samples 1 and 2 are of WSe$_2$ grown directly onto hBN, both 5.0 mm x 5.0 mm in size. Sample 3 is of MoSe$_2$-WS$_2$ on SiO$_2$/Si, 11.2 mm x 5.8 mm in size, where MoSe$_2$ was grown on c-plane sapphire and then transferred, and the WS$_2$ was grown directly onto the SiO$_2$/Si. The details of how these samples were grown can be found in Methods. The samples of WSe$_2$ were manually mapped in PL imaging mode at 20x magnification, in a similar way as described above for sample 3. The image processing tools were applied to each frame and the shape of every identified object was analysed, and an orientation measured, when possible, via the method introduced earlier. The histograms in Fig. \ref{fig:d}. (b), (d) and (f) display the results of these measurements. 

We further visualise and support our findings on the single-crystal domain orientation presented in Fig. \ref{fig:d}. (b), (d) and (f) by using Fast Fourier transforms (FFTs) analysis of the microscopy images. An FFT represents an image as a collection of two-dimensional sinusoidal waves of varying wavelength. A commonly orientated boundary, such as one side of a repeated shape in an epitaxial sample, would be represented by a line running at an angle perpendicular to the boundary edge from the centre of the FFT, assuming no pattern in nucleation position or size. For a triangular morphology and two dominant directions of alignment we thus would expect six lines extending from the centre, whereas for four dominant orientations, twelve lines should be observed. Each of the FFT graphs shown in Fig. \ref{fig:d} represent the entirety of each sample. These images were created by superimposing individual FFTs of the binary version of each frame.

An example image of sample 1 can be seen in Fig. \ref{fig:d}. (a), and on inspection no preferred orientation of individual islands can be seen. This is in agreement with the FFT, inset to Fig. \ref{fig:d}. (a), where there are no obvious lines extending radially from the centre, instead a circular shape is seen which is characteristic of randomly orientated islands. Across sample 1, an orientation could be measured for 1,419 objects from a possible 5,272 that were found by the program. Thus, 26.9\% of objects could be measured which is 5.65\% of all monolayer material by area. The low percentage of monolayer area where orientation is measurable is due to many different types of growth observed in this sample, including multi-pointed stars \cite{wang2014shape}, more irregular shapes and monolayer sheets, all composed of multiple crystal domains, which tend to be much larger objects than a single triangular island. The image shown in Fig. \ref{fig:d}. (a) is an example of the areas which can be most effectively analysed. The histogram of orientation in Fig. \ref{fig:d}. (b) shows a bi-modal distribution with peaks centred around 35$^o$ and 90$^o$, which is close to the 60$^o$ split expected from a sample with two directions of alignment at 180$^o$ to one another and a triangular morphology with three-fold symmetry. The two distributions are very broad suggesting only a small degree of alignment to the hBN substrate. 


The orientation distribution expected from typical epitaxial growth measured for sample 2 of WSe$_2$ on hBN is seen clearly in Fig. \ref{fig:d}. (d). An example image of this highly aligned sample is seen in (c), along with an inset FFT. The orientation distribution in Fig \ref{fig:d}. (d) shows two peaks centred at 37$^o$ and 95$^o$, with an almost equal number of islands contained in each. The degeneracy in number of islands at each orientation is a well documented characteristic of epitaxial growth \cite{yan2015direct,yu2017precisely,dumcenco2015large}. The narrowness of the peaks display how closely aligned the islands are to the crystal lattice of hBN with FWHMs of 9$^o$ and 7$^o$ for the 37$^o$ and 95$^o$ peaks respectively. The FFT shows six lines emerging out from the centre as expected. This sample was identical in size to sample 1, and 5,576 isolated single domain islands were measured from a possible 14,033 objects, 39.7\% by number but only 9.90\% by area. Similarly to sample 1, sample 2 contains large monolayer sheets, present on nearly a third of the sample, and irregularly shaped islands whose orientation cannot be measured through this method. However, when triangular, objects were a more consistent equilateral shape than in sample 1, making them easier to identify and measure. The image seen in Fig. \ref{fig:d}. (c) shows an area where the orientation can be seen clearly, but it is not representative of the entire sample.

The final orientation histogram, Fig. \ref{fig:d}. (f) shows the same analysis applied to sample 3 of MoSe$_2$-WS$_2$ described in Fig. 2-3. The WS$_2$ islands are low in number and only 263 of the 2,573 identified could be measured, 10.2\% of objects by number and 11.1\% by area. This small number of objects being measured is due to many of the isolated islands not being triangular, and of those that are, many have their PL quenched by the formation of heterostructures with MoSe$_2$ -- affecting their observed shape. The flat distribution suggests a lack of epitaxial growth for this material, although the relatively small sample size means any conclusions drawn are unlikely to be reliable. The MoSe$_2$ islands show a more ordered behaviour. Of the 97,672 isolated MoSe$_2$ objects identified, 27,516 could be measured by the orientation analysis program -- 28.2\% of objects by number and 23.2\% of the material by area. The much higher percentage of monolayer area that could be measured, compared to previous two samples, is due to this sample containing no areas of continuous monolayer sheets. However, irregular shapes from merged domains are still common. The image in Fig. \ref{fig:d}. (e) is a good example of the average growth across the covered sections of the sample. By eye the MoSe$_2$ islands appear to be randomly orientated, but the distribution shows four peaks spaced roughly 30$^o$ apart at 12.5$^o$, 42.0$^o$, 73.0$^o$ and 103$^o$. These results can be reconciled with the synthesis of the materials, as WS$_2$ was grown on SiO$_2$/Si and MoSe$_2$ on the hexagonal crystal structure of c-plane sapphire. These four preferential directions have been observed previously on c-plane sapphire across a 220 $\mu$m x 220 $\mu$m area \cite{dumcenco2015large}, where two of the orientations -- 60$^o$ apart -- were dominant, shown by 91.5\% of the islands, while the other two, 30$^o$ from the first were only seen in 6\%. Here, although we see a more dominant peak at 12.5$^o$, the four clear peaks highlight the existence of four significant preferential directions of growth. It appears that the strictness of alignment is lower for this sample than for sample 2 and FWHMs cannot be found accurately, but the pattern can still be observed in the FFT, inset to Fig. \ref{fig:d}. (e), with twelve lines extending from the centre.

\noindent\textbf{Discussion.}
\\For the CVD growth of monolayer MoS$_2$ on mica \cite{ji2013epitaxialSEM}, WS$_2$ and MoS$_2$ on hBN \cite{okada2014direct,yan2015direct,yu2017precisely}, and WS$_2$ on graphene, two preferential directions have been seen in previous literature with triangular islands rotated at 60$^o$ relative to one another. Both two \cite{zhang2018diffusion} and four \cite{dumcenco2015large} orientations have been observed for growth on sapphire. For the latter, the second set, 30$^o$ from the first set, had been shown to have a low probability \cite{dumcenco2015large}. To further reduce grain boundaries, it has been observed that the introduction of atom vacancies in hBN acts to trap W atoms of WSe$_2$, breaking the typical degeneracy in the number of islands at either orientation \cite{zhang2019defect}. However, assessment of aligned growth has so far been limited to tens or hundreds of islands across sub 0.1 mm$^2$ areas and therefore conclusions of large area epitaxial arrangement are hard to make considering the relatively unpredictable nature of CVD growth. Although previous work demonstrates that some degree of control is possible in CVD growth, having a means to assess the strictness of the alignment across a whole sample would allow growers to harness this control to reduce the density of grain boundaries in monolayer sheets, potentially improving electrical and optical properties of TMD films. Using image processing, we measure the strictness of epitaxial growth across a 5.0 mm x 5.0 mm WSe$_2$/hBN sample, where alignment was obvious on inspection, and uncover four prevalent orientations in MoSe$_2$ initially grown on c-plane sapphire, across a 11.2 mm x 5.8 mm MoSe$_2$-WS$_2$ sample on SiO$_2$/Si. 

With thousands of objects analysed, a statistical approach to assessing the quality of an entire substrate can be taken, as well as the ability to focus on the properties of individual islands such as, size, orientation and quality of PL emission. The analysis holds vast amounts of information, providing a deep understanding of the characteristics of a sample and the opportunity to quantitatively compare samples grown under different conditions.

There are limitations which must be noted. The merging of single domain islands causes data on epitaxially grown material to be discarded, along with a possible overestimation of size measurements and an underestimation of nucleation density. The effects on size and density can be countered with a further image processing technique called segmentation, although the artificial borders created during this process are not always accurate and it tends to overcompensate for the problem. On the other hand, for orientation, in most cases a large enough number of objects can be detected across a substrate to give a representative measurement of the level of epitaxial growth. Any technique that relies on PL imaging will encounter similar problems and a more complex method would be necessary to extract orientation and grain boundaries from merged domains. As can be seen in the Supplementary Note II, the degree of epitaxial growth of objects can still be assessed in samples of 2D materials that do not emit PL at room temperature using DF imaging, providing the correct morphology is present.

Image analysis will likely become the norm as progression in the field leads to inevitable commercial production of devices that require a consistent quality of TMD monolayer, without time for constant manual inspection. The utility of the automated analysis developed in this work comes in its ability to characterise the quality of grown material automatically and unearth patterns that would be missed by manual inspection, such as the four preferential directions found in material grown on c-plane sapphire, Fig. \ref{fig:d}. (f), and the quantitative level of homogeneity in MoSe$_2$ island size across the same sample, Fig. \ref{fig:b}. (d). 

To conclude, a new automated analysis has been developed capable of measuring the characteristics of monolayer islands across tens of mm$^2$ areas of CVD grown TMD semiconductors, from images that map the sample surface. The analysis uses PL imaging techniques and can be applied to various combinations of semiconducting TMD monolayer and substrate which emit PL at room temperature. Functions from MATLAB's image processing tool box are used to measure island size and density, as well as relative average value, and homogeneity of, PL emission. Importantly, for each isolated monolayer island across a sample we measure orientation, and obtain data on the presence and degree of epitaxial growth. An FFT analysis has been further applied to uncover any underlying patterns in the sample and to verify the results of the orientation histogram. Four preferential directions of alignment, not apparent to the naked eye, were found in MoSe$_2$ islands grown on c-plane sapphire, a characteristic that had not been seen previously to such a degree. This analysis method coupled with the automated mapping of CVD samples in PL image mode forms a fast and reliable characterisation tool. With the use of dark-field imaging the analysis of epitaxial growth can be extended to many other materials that do not emit PL. Such analysis will reduce the amount of time needed to study CVD samples or 2D materials produced by other scalable techniques, and has particular application to developing a consistent method of production for high quality 2D monolayer sheets and heterostructures. Applications in other thin film material systems, such as perovskite or organic semiconductors are also possible.

\bigskip

\textbf{METHODS}
\bigskip

\textbf{Growth of single layer WS$_2$.}
WS$_2$ was grown on a SiO$_2$ substrate via CVD. Before the growth of WS$_2$, SiO$_2$ was cleaned by a piranha solution and was spin-coated by sodium cholate which acted as a seeding promoter. WO$_3$ (99.998\%, Alfa Aesar) was dissolved in the water/ammonia solution (9:1) and 1 mL of the solution was coated on the crucible. This crucible was placed at the centre of the furnace and 100 mg of S (99.999\%, Alfa Aesar) was placed at the upstream entry of the furnace. Then, the temperature of the tube furnace was increased up to 900 $^o$C for 24 min under a steady flow of Ar gas (100 sccm) in the ambient condition. When the furnace reached 600 $^o$C the S vaporised. Then, temperature of the tube furnace was maintained at 900 $^o$C for 30 min for the WS$_2$ growth. Afterwards, the tube furnace was cooled down to room temperature under the Ar flow. 

\bigskip

\textbf{Growth of single layer MoSe$_2$ and transfer to SiO$_2$/Si.}
MoSe$_2$ was grown on c-plane sapphire by CVD. Two precursors, MoO$_3$ (99.97\%, Sigma Aldrich) and Se (99.999\%, Alfa Aesar), were used for the growth. 150 mg of Se was placed at the upstream entry of the furnace and 60 mg of MoO$_3$ powder was placed at the centre of the furnace. A crucible containing MoO$_3$ was partially covered by a SiO$_2$/Si wafer to reduce intense evaporation of the precursor. The sapphire substrate was located next to the crucible that contained MoO$_3$. Before the tube furnace was heated, the tube was evacuated for 30 min and filled with the Ar gas achieving ambient pressure. The temperature of the furnace was increased up to 600 $^o$C for 18 min under a steady flow of Ar gas (60 sccm) and H$_2$ gas (12 sccm). When the furnace reached 600 $^o$C, Se was vaporised by heating the upstream entry of the tube up to 270 $^o$C using a heating belt. Finally, temperature of the tube furnace was increased to 700 $^o$C and maintained for 1 hr for the MoSe$_2$ growth. Afterwards, the tube furnace was cooled down to room temperature while the Ar flow was maintained without H$_2$. To transfer MoSe$_2$ on top of the SiO$_2$/Si substrate containing WS$_2$, polystyrene was used to maintain the sample quality instead of poly(methyl methacrylate) that has been widely used previously \cite{gurarslan2014surface}.

\bigskip

\textbf{Growth of single layer WSe$_2$ on hBN.} 
To fabricate WSe$_2$ on hBN, multilayer hBN was initially grown on c-plane sapphire \cite{jang2016wafer}. WO$_3$ (99.998\%, Alfa Aesar) and Se (99.999\%, Alfa Aesar), were used for the WSe$_2$ growth. 300 mg  of Se was placed at the upstream entry of the furnace and 120 mg of WO$_3$ powder was placed at the centre of the furnace. To reduce the influence of humidity, a small amount of NaCl was mixed with WO$_3$ powder. The multilayer hBN on sapphire substrate was positioned next to the crucible containing WO$_3$. Before the tube furnace was heated, the tube was evacuated for more than 30 min. Then, the temperature of the tube furnace was increased to 800 $^o$C for 24 min under a steady flow of Ar gas (120 sccm) and H$_2$ gas (20 sccm). When the furnace reached 800 $^o$C, the Se was vaporised by heating the upstream entry of the tube to 270 $^o$C using a heating belt. Finally, temperature of the tube furnace was increased to 870 $^o$C and maintained for 1 hour for the WSe$_2$ growth. Afterwards, the tube furnace was cooled down to room temperature under Ar flow. 

\bigskip

\textbf{PL imaging.} 
The PL images analysed, such as Fig. \ref{fig:a}. (d), were taken with an adapted industrial microscope (LV150N, Nikon). A 550 nm short-pass filter is positioned in the path of the white light from the illumination source and a 600 nm long pass filter is placed before the camera (DS-Vi1, Nikon) to isolate PL emission. A further 550 long pass dichroic mirror is applied to direct the excitation light and collected PL emission. A detailed description of the method can be found in Ref. \cite{alexeev2017imaging}. Using this technique the samples discussed in reference to Fig. \ref{fig:a} through \ref{fig:d} were mapped manually at 20x magnification. The images of WSe$_2$ on hBN were taken with 4 s exposure time and an analogue gain of 5.6x, those of MoSe$_2$-WS$_2$ on SiO$_2$/Si were taken with 10 s and 9.6x, respectively. Each image was 373 $\mu$m x 497 $\mu$m in area.

\bigskip

\textbf{Image processing.}
Images were analysed in MATLAB using functions from the image processing toolbox \cite{matlabImage}. The colour thresholding application was used to isolate monolayer material in a PL image and was applied to each combination of  monolayer and substrate. The `regionprops' function was used for the analysis of size, island density and pixel value with minimal custom code. The program that measured the orientation of equilateral triangles was developed specifically for the analysis of epitaxial growth. It took the program 43 seconds to analyse  the orientation of the images for sample 1 containing 5,272 objects, 69 seconds to analyse that of sample 2 of 14,033 objects, and 1 hour 8 minutes and 13 seconds to analyse sample 3 containing 100,245 objects of two different colours. FFTs, representing entire substrates, were created by superimposing individual image FFTs and were subsequently artificially enhanced for clarity by transforming the pixel values using contrast-limited adaptive histogram equalisation. Further details of the image processing and a more complete explanation of the analysis can be found in the Supplementary Note I.



\bigskip

\textbf{Acknowledgements}

\noindent
T.S.M., A. G., E.M.A. and A. I. T. acknowledge funding by EPSRC (EP/P026850/1). S. R. and A. I. T. acknowledge support from the SURE scheme of the University of Sheffield. This work was supported by the research fund (NRF-2017R1E1A1A01074493) by the Ministry of Science and ICT, Korea.

\bigskip

\textbf{Author contributions}
The samples were grown by S. A., A. J. and H. S. S. S. R. manually mapped the sample of MoSe$_2$-WS$_2$ and T.S.M. mapped those of WSe$_2$. T.S.M., A.G. and E.M.A. developed the MATLAB script for the computer analysis. T.S.M. analysed all images and produced data for the manuscript. T.S.M., A.G. and A.I.T. wrote the manuscript with contributions from all co-authors. A.I.T. conceived and supervised the project.

\bigskip

\textbf{Competing interests}

The authors declare no competing interests


\newpage

\renewcommand{\thefigure}{S\arabic{figure}}

\setcounter{figure}{0}

\noindent
\textbf{\Large{Supplementary Information for ``Large Area Characterisation of Chemical Vapour Deposition Grown Monolayer Transition Metal Dichalcogenide Through Optical Imaging'' }}

\textbf{Supplementary note I.}
\\\textbf{Image processing in MATLAB \& orientation measurement.} The image processing methods required to analyse the physical and optical properties of CVD samples can be found in MATLAB's image processing tool box \cite{matlabImage}. A calibration image of a sample which contains all materials to be analysed is loaded into the colour thresholding application and each material isolated manually using the colour channels whose values are then exported into the main program. Alternatively, if only one material is present, a simpler form of thresholding can be applied that isolates any sections with a pixel intensity greater than 1.5 times the mean pixel value of the calibration image. Results obtained using the two thresholding methods for an aligned WSe$_2$ sample grown on hBN were identical and run times comparable. 
\\Once either method has been applied to create a binary image (BW), where white areas are monolayers and black areas represent substrate or thicker material, the function `imclearborder(BW)' is called to remove objects that are not completely contained in the image. This causes some islands to be missed by the program, but is justified by the fact that false negatives have less of a detrimental impact on results than the false positives that are created when objects have been altered by the limited field of view of the camera. This loss of data matters little for large samples containing thousands of objects. The function `imfill(BW, 8, `holes')' is then used to fill in any holes in an object surrounded by pixels with at least 8 connections and `bwareaopen(BW, P, 8)' is applied to remove objects with `P' number of pixels with the same connectivity. Both processes improve the analysis, the former allows the binary image to be used as a mask and the latter is necessary to remove noise and reflections -- reasonably common features in PL images. Samples 1 and 2 in the main paper contain large areas of monolayer sheets which need to be analysed despite it not being possible to extract any information on epitaxial arrangement. These areas will almost always be in contact with the border of an image. To ensure these are not removed from the analysis, objects of 384,000 pixels or greater in size (one fifth of the image size) that do touch the image border are not removed from the image, but an orientation measurement is not attempted.

\renewcommand{\baselinestretch}{1.2}

\begin{figure*} [b!]
	\centering
	\includegraphics[width=1\linewidth]{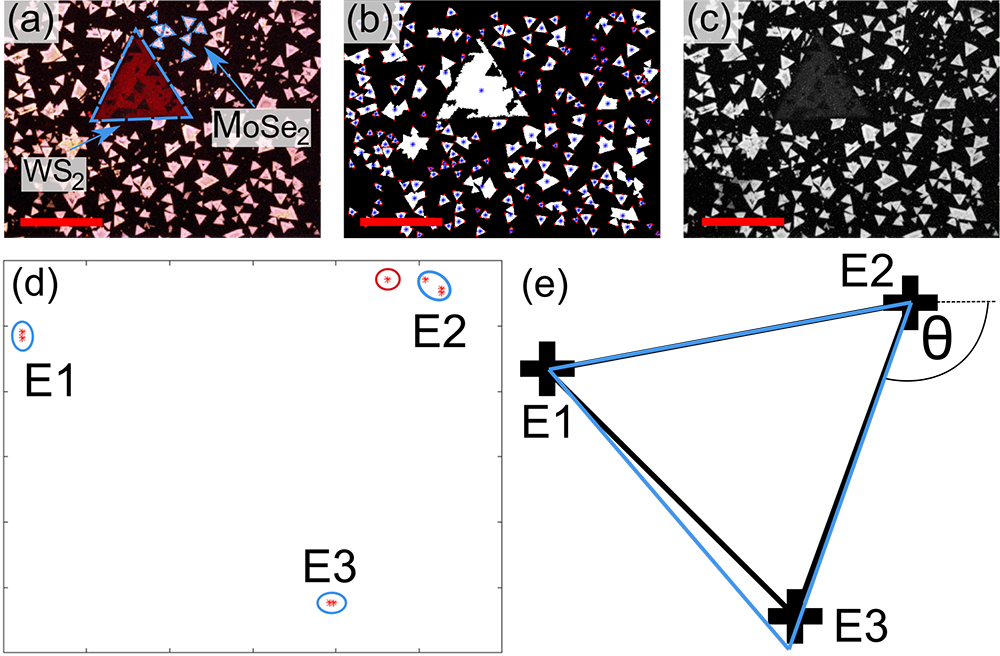} 
	\caption{\textbf{Image processing and program logic of the orientation analysis.} (a) Shows a PL image at 50x magnification of MoSe$_2$-WS$_2$ on SiO$_2$/Si, with the two materials clearly distinguishable by colour, as is necessary for such mixed material images to be convert to binary form. The binary image created from the PL version can be seen in (b) with a centroid for each object highlighted by blue asterisk and the extrema by a red marker. (c) Is the grayscale image of the same area. All red scale bars are 50 $\mu$m. The logic of the program when trying to determine if an object is the desired triangular shape is represented in (d), the red asterisks mark the positions of the extrema from a single object. Those extrema that will be averaged based on proximity are circled in blue and the one which will be discarded is circled in red. (e) Demonstrates how the program finds the orientation from the combined extrema, E1-3, by fitting an equilateral triangle, shown in blue, to each side.
	\label{fig:e}}
\end{figure*}

\noindent A grayscale image is also created from the full colour version, where each pixel is represented as a value from 0, for black, to 255, for white, with numbers between being tones of grey (for 8 bit images). The value of a pixel after this conversion depends upon the channels of the image format that define the pixels initial colour. Each colour channel is weighted differently and an average taken across all, in a standard conversion from full colour to grayscale image. Because of this, purely red, green, or blue pixels will have small pixels values compared to composite colours and those of two different colours can not be compared in terms of brightness. The effect of this weighted conversion is seen by comparing Fig. \ref{fig:e}. (a) and (c), where the large red WS$_2$ flake is barely distinguishable from the substrate, while those of the light pink MoSe$_2$ are clear. Fig. \ref{fig:e}. (a) and (b) demonstrate the effects of processing a false colour PL image (a) to become a binary image (b). This binary image is then applied as a mask to the grayscale (c) of the original colour image through the function `regionprops(BW, grayscale, property1, property2,...)', to select just those areas that meet the criteria of the binary image.
\\The properties taken from the binary image are size of isolated objects, as well as the number of objects contained in an image, and the positions of the extrema for each. The grayscale image is necessary as it contains information on the pixel values and is used to find the properties of mean pixel value for each object and the standard deviation (STD) from that mean across the face of those objects. These provide a relative measure of average PL emitted by a monolayer island and the homogeneity of that emission across its surface, respectively. 
\\The extrema positions for each object are used to determine if an object is a triangle and, if so, measure the orientation. As a pre-built function of the image processing toolbox, `regionprops' gives a set of 8 extrema for each object. If any two or more extrema are within a certain distance of each other, determined by the overall size of the object, their positions are averaged together to create a single point. A representation of this step is seen in Fig. \ref{fig:e}. (d), where the red extrema that will be combined are circled in blue and those that will be discarded are circled in red. Any triangular shape will be left with exactly three extrema and an object found to have any other number remaining is discarded from the data set. In order to find the orientation, an equilateral triangle is fitted to each combination of two remaining extrema points and the smallest angle relative to an imaginary horizontal axis is found for each. A visual representation of programs logic during this step is shown in (e), with the equilateral triangle being fitted to combined extrema E1 and E2. The average of the three orientation values is taken and the range is used to calculate an uncertainty, with those objects closest to the expected equilateral triangle shape having the lowest uncertainty. Therefore, by discarding all objects with a percentage uncertainty of more than 30\% of the mean value for the data set, only islands with the expected morphology are selected.

\begin{figure*} [b!]
	\centering
	\includegraphics[width=1\linewidth]{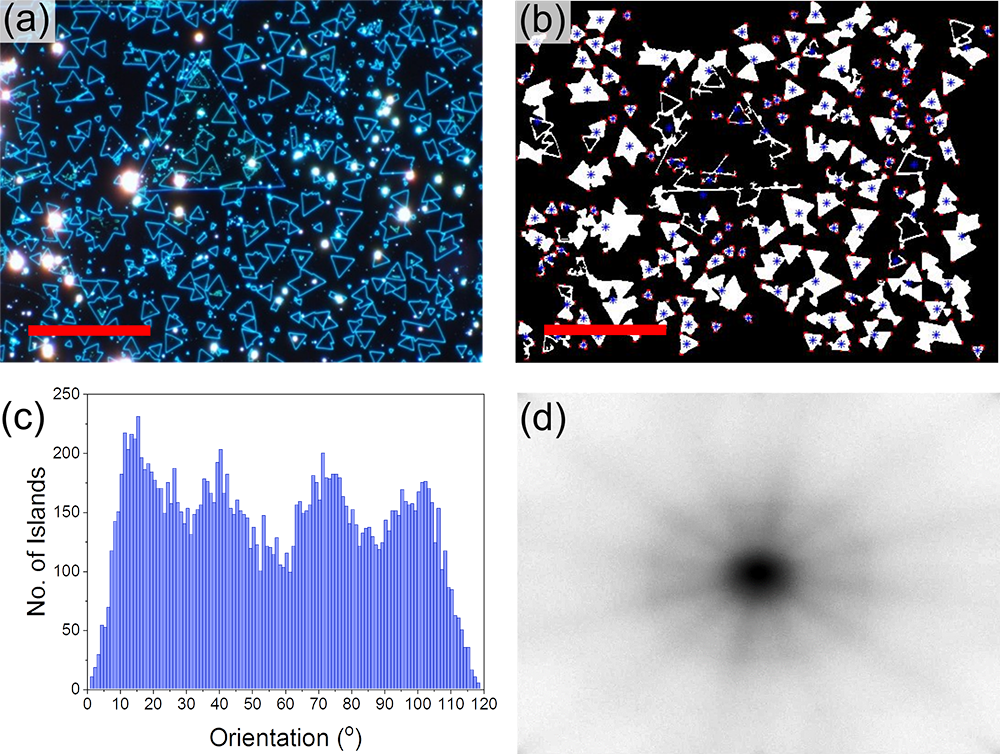} 
	\caption{\textbf{Orientation analysis from DF images.} (a) Is a 50x magnification DF image of MoSe$_2$ and WS$_2$ islands on SiO$_2$/Si. The binary image of the DF image after the processing tools are applied is seen in (b). These can be contrasted with Fig. \ref{fig:e}. (a) and (b), respectively, for a comparison between PL and DF imaging modes. (c) Is the orientation distribution obtained when the analysis is applied to DF images that completely map the MoSe$_2$-WS$_2$ on SiO$_2$/Si. 
	\label{fig:f}}
\end{figure*}

\textbf{Supplementary note II.}
\\\textbf{Orientation analysis of dark-field (DF) images.} PL imaging provides the highest contrast between material and substrate, as well as wavelength dependent colours for different materials. However, there are many 2D materials that do not emit PL at room temperature but can still be grown via CVD, and it is beneficial to be able to measure the degree of epitaxial arrangement in such materials. DF imaging, Fig. \ref{fig:f}. (a), is the natural choice for image analysis of non emitting materials, as the contrast between the edges of material and the substrate is much higher than in BF. Very similar processing tools are applied to the DF as were to the PL images. Colour thresholding is applied to isolate areas of dark blue in the image, to leave just the edges of islands. Boarders are cleared, small objects are removed and completely surrounded areas are filled in, to leave a binary image similar to that seen in Fig. \ref{fig:f}. (b). When compared to the binary image seen in Fig. \ref{fig:e}. (b) it is apparent that the imaging modes do not collect exactly the same features. In the DF version, dust, which appears as the bright white dots in Fig. \ref{fig:f}. (a), masks features of islands. Those islands in close proximity to such contamination are warped in shape or missing from the binary image. The large WS$_2$ flake in the centre of the image is not picked up as an object in the binary image, predominantly due to a dust particle being present near to bottom left vertex. Even if the island was found, it would not be possible to distinguish it from the MoSe$_2$ islands, without already knowing that WS$_2$ islands are much larger.

\noindent The orientation results obtained when the analysis is applied to a collection of DF images which completely map the sample of MoSe2$_2$-WS$_2$ on SiO$_2$/Si are seen in Fig. \ref{fig:f}. (c). The DF images were taken in exactly the same position as those of PL, and the histogram resembles that of the MoSe$_2$ orientation results seen in Fig. 4. (f) of the main paper, with four peaks displayed. Clear differences between the DF and PL results are the reduction in size of the peak at 12.5$^o$ and the smoothing out of all peaks in general. As there is no way to distinguish the MoSe$_2$ material from the WS$_2$, the histogram is a superposition of both orientations, promoting the smoothing of the MoSe$_2$ features. Using DF imaging 16,137 objects could be measured from 79,208 identified, compared to 27,779 from 100,245 identified in the PL images.  Given the large impact of the contaminants on the sample these results seem promising for the use of DF imaging to assess epitaxial growth across samples. 
\\The other property DF imaging can be used for, island size, is more problematic. As only the edges of objects are imaged in DF, a slight change in colour leads to a break in the object edge causing it not to be filled in. The measured size will be considerably less and characterisation results will be skewed. 
\\Every image that mapped the sample was converted into binary form and a fast Fourier transform (FFT) applied. Each image FFT is then added together to create one which represents the orientation results for the entire sample. The full sample FFT is shown in Fig. \ref{fig:f}. (d) and resembles that seen in the inset to Fig. 4. (e) of the main paper, with twelve lines extending radially from the centre. The lines appear weaker here which is surprising as those triangles which could not be filled in have a second set of boundaries that should enhance the lines. Conversely, the contaminants do lead to objects being removed completely. These preliminary results based on DF imaging are promising and with more optimisation this method can be applied to a broad class of 2D materials that do not emit PL.


\end{document}